# Structural transition, electric transport, and electronic structures in the compressed trilayer nickelate $La_4Ni_3O_{10}$

Jingyuan Li[1,*], Cui-Qun Chen[1,*], Chaoxin Huang[1], Yifeng Han[2], Mengwu Huo[1], Xing Huang[1], Peiyue Ma[1], Zhengyang Qiu[1], Junfeng Chen[1], Xunwu Hu[1], Lan Chen[1], Tao Xie[1], Bing Shen[1], Hualei Sun[3,#], Dao-Xin Yao[1,§], and Meng Wang[1,†]

[1] *Center for Neutron Science and Technology, Guangdong Provincial Key Laboratory of Magnetoelectric Physics and Devices, School of Physics, Sun Yat-Sen University, Guangzhou, Guangdong 510275, China*

[2] *Center for Materials of the Universe, School of Molecular Sciences, Arizona State University, Tempe, AZ 85287, USA*

[3] *School of Sciences, Sun Yat-Sen University, Shenzhen, Guangdong 518107, China*

*# sunhlei@mail.sysu.edu.cn*

*§ yaodaox@mail.sysu.edu.cn*

*† wangmeng5@mail.sysu.edu.cn*

**Abstract**

Atomic structure and electronic band structure are fundamental properties for understanding the mechanism of superconductivity. Motivated by the discovery of pressure-induced high-temperature superconductivity at 80 K in the bilayer Ruddlesden-Popper nickelate $La_3Ni_2O_7$, the atomic structure and electronic band structure of the trilayer nickelate $La_4Ni_3O_{10}$ under pressure up to 44.3 GPa are investigated. A structural transition from the monoclinic $P2_1/a$ space group to the tetragonal $I4/mmm$ around 12.6~13.4 GPa is identified, accompanying with a drop of resistance below 7 K. Density functional theory calculations suggest that the bonding state of Ni $3d_{z^2}$ orbital rises and crosses the Fermi level at high pressures, which may give rise to possible superconductivity observed in resistance under pressure in $La_4Ni_3O_{10}$. The trilayer nickelate $La_4Ni_3O_{10}$ shows some similarities with the bilayer $La_3Ni_2O_7$ and has unique properties, providing a new platform to investigate the underlying mechanism of superconductivity in nickelates.

# 1 Introduction

Exploration of new superconductors has been one of the most important frontiers in condensed matter physics since the discovery of superconductivity in mercury in 1911. After discovering high-transition-temperature (HT$_c$) superconductivity in cuprates in 1986, many efforts have been devoted to elucidating the mechanism of unconventional superconductivity and searching for superconductivity in oxide compounds. Given that Ni$^+$ has the same 3$d^9$ electron configuration as Cu$^{2+}$, nickelates were predicted to host HT$_c$ far back in 1999 [1]. However, the first experimental realization of superconductivity in nickelates was in the thin film samples of reduced Ruddlesden-Popper (RP) phase Nd$_{0.8}$Sr$_{0.2}$NiO$_2$ in 2019 [2]. The superconducting transition temperature ($T_c$) was 9-15 K, which is much lower than that of the isostructural cuprate superconductors. The low $T_c$ in Nd$_{0.8}$Sr$_{0.2}$NiO$_2$ can be ascribed to the absence of hybridization between the Ni 3$d$ and O 2$p$ orbitals because the O 2$p$ orbitals are far below the Fermi level [3, 4].

Recently, Sun *et al* reported the discovery of superconductivity in the pressurized bilayer Ruddlesden-Popper (RP) phase La$_3$Ni$_2$O$_7$ with the maximum $T_c$ at 80 K [5]. This discovery established nickelates as a new HT$_c$ superconducting family under pressure and thus reawakened intense research on nickelates [6-26]. The valence state of Ni ions in La$_3$Ni$_2$O$_7$ is a mixture of +2 and +3, yielding an averaged state of Ni$^{2.5+}$ (3$d^{7.5+}$) state [27]. In coincidence with the emergence of the pressure-induced superconductivity, La$_3$Ni$_2$O$_7$ undergoes a structural transition from the *Amam* to the *Fmmm* space group. In the superconducting state, a structural transition from the orthorhombic *Fmmm* space group to the tetragonal *I*4/*mmm* space group was suggested [24]. The interlayer Ni-O-Ni bond angle approaches 180° in the *Fmmm* and *I*4/*mmm* space groups. It was suggested that the formation of 180° Ni-O-Ni angle strengthens the hybridization between the Ni 3$d_{z2}$ and O 2$p_z$ orbital and finally induces superconductivity [28]. In cuprates, the hybridization between the 3$d_{x2-y2}$ and 2$p_{x/y}$ orbitals can cause a strong superexchange interaction that assists the mediation of electron pairing to form the HT$_c$ superconductivity.

It is of high interest to testify whether the HT$_c$ superconductivity could be realized in the trilayer RP phase La$_4$Ni$_3$O$_{10}$. In terms of cuprates, the trilayer systems hold the highest $T_c$ [29,30]. At ambient pressure, La$_4$Ni$_3$O$_{10}$ is metallic with intertwined spin density wave and charge density wave transitions at ~140 K [31-33]. The averaged valence state of Ni ions is Ni$^{2.67+}$ (3$d^{7.33+}$). Moreover, according to angular resolved photoelectron spectroscopy (ARPES) measurements, a flat γ band with strong 3$d_{z2}$ character lies close to the Fermi level [34]. By comparison with the electronic band structure of La$_3$Ni$_2$O$_7$, potential superconductivity is desirable in La$_4$Ni$_3$O$_{10}$ under pressure. Recently, clear drops in resistance of La$_4$Ni$_3$O$_{10}$ under pressure at about 20-30 K were observed [35-38], indicating signatures of superconductivity. However, detailed studies on the structure of La$_4$Ni$_3$O$_{10}$ and electronic band structure under pressure are still absent.

In this work, we report a comprehensive experimental and theoretical study on La$_4$Ni$_3$O$_{10}$ single crystals under pressure. The single crystal samples were grown by a high oxygen pressure optical floating zone furnace. Synchrotron X-ray diffraction studies reveal a structural transition from the monoclinic *P*2$_1$/*a* space group to the tetragonal *I*4/*mmm* space group at about 12.6-13.4 GPa. Our density functional theory (DFT) calculations of the band structures show the 3$d_{z^2}$ bonding band of Ni ions cross the Fermi level under high pressure, which mimics the pressure-induced 3$d_{z^2}$ σ-bond metallization in La$_3$Ni$_2$O$_7$. In addition, investigations on the high-pressure transport properties of La$_4$Ni$_3$O$_{10}$ show a weak drop in resistance at 7 K using the diamond anvil cell (DAC) method. Thus, if the superconductivity in the tetragonal phase of La$_4$Ni$_3$O$_{10}$ indeed exists, it would be also sensitive to the content of oxygen and the homogeneity of pressure.

## 2 Experimental and calculation methods

La$_4$Ni$_3$O$_{10}$ precursor polycrystalline rods were synthesized by solid-state reaction method from La$_2$O$_3$ and NiO at 1400°C. High-quality single crystals of La$_4$Ni$_3$O$_{10}$ were grown in a vertical optical-image floating zone furnace at an oxygen pressure of 20 bar and a 5-kW Xenon arc lamp (100-bar Model HKZ, SciDre). The structure of La$_4$Ni$_3$O$_{10}$ was confirmed by fine powders ground from the single crystal samples. Magnetic susceptibility and resistivity were performed on a physical property measurement system (PPMS, Quantum Design).

The *in situ* high-pressure synchrotron X-ray diffraction (XRD) patterns of powder La$_4$Ni$_3$O$_{10}$ were collected at 300 K with a wavelength of 0.6199 Å on Beijing Synchrotron Radiation Facility, Institute of High Energy Physics, Chinese Academy of Sciences (BSRF, IHEP, CAS). Details of the sample preparation for the high-pressure XRD measurements are identical to our measurements on La$_3$Ni$_2$O$_7$, which have been described elsewhere [5].

High-pressure electric transport measurements of La$_4$Ni$_3$O$_{10}$ single crystals were carried out using a miniature DAC made from a Be–Cu alloy. Diamond anvils with a 300-μm culet were used, and the corresponding sample chamber with a diameter of 100-μm was made in an insulating gasket achieved by cubic boron nitride and epoxy mixture. The pressure was also calibrated by measuring the shift of the fluorescence wavelength of the ruby sphere, which was loaded in the sample chamber. The four-probe van der Pauw method was adopted for these measurements. No pressure-transmitting medium was adopted in the measurements.

Density functional theory (DFT) calculations were performed using the Vienna *ab initio* Simulation Package (VASP) [39]. To obtain the same Fermi surface of La$_4$Ni$_3$O$_{10}$ at ambient pressure with the ARPES measurements [34], the density functional was estimated using the local density approximation (LDA) exchange-correlation potential, and the Columb interaction was involved by an effective Hubbard U as 0.5 eV [40]. We adopted the projector augmented wave with a plane-wave cutoff energy of 600 eV. An 18×18×4 k-point mesh was employed for self-consistent and Fermi surface calculation. The experimentally measured lattice constants were used in our calculations and the positions of all atoms were fully relaxed until the force on each atom was less than 0.001 eV/Å. The energy convergence criterion was set at $10^{-7}$ eV for the electronic self-consistent loop.

## 3 Results

### 3.1 Properties at ambient pressure

We performed characterizations on the structure, electronic transport, and magnetic susceptibility of the $La_4Ni_3O_{10}$ single crystals under ambient pressure. The crystal structure of $La_4Ni_3O_{10}$ is shown in Figure 1(a). Though there are different structural phases of $La_4Ni_3O_{10}$ at room temperature [41-44], as the powder XRD patterns and its Rietveld refinement results show, the structure can be properly fitted by the monoclinic $P2_1/a$ (Z=2) space group. Good matching between the observed peaks and Rietveld refinements indicates the purity and quality of the sample. The refined unit cell parameters are $a$ = 5.4164(2) Å, $b$ = 5.4675(2) Å, $c$ = 14.2279(3) Å, and $β$=100.752(3)°, in good agreement with previous high-resolution synchrotron XRD measurement [41].

Figure 1(b) shows the temperature dependence of the resistance and zero-field cooling magnetic susceptibility of $La_4Ni_3O_{10}$ single crystal at ambient pressure. The in-plane resistance drops with decreasing temperature rapidly, exhibiting a metallic behavior. An anomaly appears at around $T^* \sim$ 136.5 K, coincident with the drop in the out-of-plane magnetic susceptibility [31-33]. The transition-like behaviors in resistance and susceptibility can be ascribed to the emergence of intertwined charge and spin density waves of the $P2_1/a$ phase [33].

### 3.2 Structural transition under pressure

Because of the importance of the high-pressure structure study, we conducted *in situ* high-pressure synchrotron XRD measurements on $La_4Ni_3O_{10}$ up to 44.3 GPa at room temperature. Figure 2(a) displays the high-pressure XRD patterns of $La_4Ni_3O_{10}$ under pressure from 1.8 to 44.3 GPa. As can be seen in Figure 2(c), the structure at low pressure can be properly fitted by the monotonic space group $P2_1/a$, consistent with the space group at ambient pressure. Figure 2(b) shows the enlarged details of the evolution of (1 1 3), (0 2 0), and (2 0 -1) peaks. Clear evidence shows that the (0 2 0) and (2 0 -1) peaks gradually merge as the pressure approaches 12.6 GPa, indicating a structural phase transition to the tetragonal structure.

The lattice parameters under different pressures obtained from Rietveld refinements are shown in Figure 2(e) and Figure 2(f). Specific information of the lattice parameters at selected pressures obtained from Rietveld refinements are displayed in TABLE S1 in the supplementary information. The corresponding refined and theoretically optimized atomic positions are displayed in TABLE S2-S13 in the supplementary information. While the lattice parameters are compressed by increasing pressure, the lattice constant

*b* shrinks more quickly than *a*. They tend to be closer as the pressure approaches 12.6 GPa. the lattice constant *c* undergoes a sudden drop during the structural phase transition. These results suggest the low-pressure (LP) phase turns into a tetragonal high-pressure (HP) phase with a higher symmetry. As shown in Figure 2(d), the HP phase structure can be properly fitted by the *I4/mmm* space group. In this case, the (0 2 0) and (2 0 -1) peaks merge into the (1 1 0) peak of the *I4/mmm* space group. In the notation of the HP phase, the in-plane lattice constant is $1/\sqrt{2}$ of that in the LP phase. For the comparisons in Figure 2(e) and 2(f), the factor of different unit cells has been included. The evolution of *V* as a function of pressure can be fitted by the Birch-Murnaghan equation [45]. The bulk modulus $B_0$ is fitted to be $B_{0,\ LP}$ = 188 ± 1 GPa for the LP phase and $B_{0,\ HP}$ =194 ± 3 GPa for the HP phase, using the data from 1.8-11.5 and 12.6-44.3 GPa, respectively. The derivative bulk modulus B' is fitted to be $B'_{LP}$ = 4.3 ± 0.08 for the LP phase and $B'_{HP}$ = 3.99 ± 0.05 for the HP phase. The distinct $B_0$ and B' further prove the existence of a structural transition. Figure 2(g) depicts the 164.8° Ni-O-Ni angle between the adjacent octahedra layers forced to 180.0° in the HP phase, reminiscent of the bilayer $La_3Ni_2O_{7-\delta}$ [5]. The uneven $NiO_2$ planes become flattened in the HP phase.

### 3.3 High-pressure transport properties

Previous high-pressure electric transport measurements reveal various onset pressures and transition temperatures for the signatures of superconductivity on powder [36,38] and single-crystal samples [35,37]. Figure 3 shows the high-pressure transport results under various pressures in our single crystal samples. The resistance at 1.1 GPa exhibits a semiconductor-like behavior in the whole temperature range. The kink indicating the intertwined charge and spin density waves at 136.5 K can be observed. A moderate drop around 7 K is noticed at 10.0 GPa, coincidence with the structural transition pressure within the error of pressure. The drop becomes more prominent as increasing pressure and persists to 23.0 GPa, which is the maximum pressure of our measurements. A magnetic field of 1 T can suppress the drop in resistance, consistent with the suppression of superconductivity by a magnetic field. The transition temperature is lower than the reports in the literature, which may be due to the various oxygen contents of the samples grown by different groups. The non-zero resistance under pressure could be ascribed to the inhomogeneity of the pressure as the studies in $La_3Ni_2O_7$ [5-9].

### 3.4 Electronic band structure calculation

To gain insight into the possible superconductivity in $La_4Ni_3O_{10}$ under pressure, we performed DFT calculation at ambient pressure and 44.3 GPa. As indicated by Figure 4(a), the band structure at ambient pressure shows a similar alignment to the previous study [34]. The $3d_{z2}$ and $3d_{x2-y2}$ orbitals are separated from the $t_{2g}$ orbitals due to the crystal field, resulting in a large proportion of $3d_{z2}$ and $3d_{x2-y2}$ orbitals around the Fermi level. Figure 4(b) shows the enlarged details of the band structure around the Fermi level. At the Γ point, a tiny gap of 27 meV separates the $3d_{z^2}$ band away from the Fermi energy while the $3d_{x^2-y^2}$ band crosses the Fermi level, forming an electron pocket ($\beta$ band) as revealed by the electron pocket centered at the Γ point in the Fermi surface as shown in Figure 4(c). Below and above the Fermi level, there are electronic bonding, non-bonding, and anti-bonding bands of the $3d_{z^2}$ states due to the interlayer σ-bond formed by the hybridization of $3d_{z^2}$ band of Ni and $2p_z$ band of O.

The Fermi level shifts downwards, and the Fermi surface changes at 44.3 GPa, which looks like hole doping for the Ni $3d_{z^2}$ orbital, as shown in Figure 4(d). The bandwidth of Ni-$3d_{z^2}$ and $3d_{x^2-y^2}$ orbitals increases. The energy span of the $3d_{x^2-y^2}$ orbital of Ni is broadened from 3.77 to 4.71 eV as the pressure increases from 0 to 44.3 GPa. Importantly, the pressure induces the intertwining of the bonding bands and non-bonding bands, which closes the tiny gap near the Γ point [Figure 4(e)]. As a result, an additional electron pocket around the center of the Brillouin zone appears, as displayed in Figure 4(f). This phenomenon resembles the behavior of pressurized $La_3Ni_2O_7$, where the bonding state is critical to the appearance of superconductivity. However, the metalized non-bonding band may compete with the bonding state and prohibit the formation of Cooper pairs in the trilayer RP phase of $La_4Ni_3O_{10}$. Therefore, the $T_c$ of the possible superconductivity in pressurized $La_4Ni_3O_{10}$, if it exists, may not be comparable to $La_3Ni_2O_7$.

## 4 Discussion and Summary

To date, neither the zero resistance nor the diamagnetic response was realized in $La_4Ni_3O_{10}$ under pressure. However, given the technical challenges of high pressure and difficulty in controlling the oxygen content of nickelates, the existence of superconductivity in $La_4Ni_3O_{10}$ under pressure is still plausible. The various behaviors of signatures of superconductivity in $La_4Ni_3O_{10}$ under pressure may be related to the inhomogeneous oxygen content that results from the conditions of sample synthesis [35-38]. The signature of superconductivity emerges at 10.0 GPa in our sample, coincident with the structural transition. According to the structural study of $La_3Ni_2O_7$ under high pressure and low temperature [24,10], the tetragonal *I4/mmm* phase may be responsible for the superconducting state. Interestingly, many cuprates [46,47] and iron pnictide/chalcogenide [48,49] superconductors exhibit the *I4/mmm* phase. Further investigations are required to uncover the relationships between the tetragonal structure and HT$_c$ superconductivity in the RP phase nickelates.

In summary, we have studied the structure of $La_4Ni_3O_{10}$ up to 44.3 GPa and calculated the electronic band structure accordingly. A tetragonal structure transition is revealed in coincidence with the appearance of the drop in resistance below 7 K. Through DFT calculations, we find that upon applying pressure, the $3d_{z^2}$ bonding band crosses the Fermi energy. The structural transition makes a significant modification on the Fermi surface, leading to the emergence of an additional electron pocket around the center of the Brillouin zone. Such a scenario resembles the $3d_{z^2}$ σ-bond metallization in $La_3Ni_2O_7$. In this sense, a possible superconductivity is desirable in pressurized $La_4Ni_3O_{10}$. However, the existence of the additional non-bonding state between the anti-bonding and bonding states may compete with the bonding state and interfere with the electron pairing in $La_4Ni_3O_{10}$ and lower the $T_c$. Our results suggest that $La_4Ni_3O_{10}$ is a promising material to explore the even-odd layer effect of the RP phase nickelates and the intrinsic association among the structure symmetry, electronic band structure, and $HT_c$ superconductivity.

## Acknowledgments

This work was supported by the National Natural Science Foundation of China (grant nos. 12174454, 12304187, U213010013, 92165204, 11974432), the Guangdong Basic and Applied Basic Research Funds (grant no. 2021B1515120015), the Guangzhou Basic and Applied Basic Research Funds (grant nos. 202201011123, 2024A04J6417), NKRDPC-2022YFA1402802, the Guangdong Provincial Key Laboratory of Magnetoelectric Physics and Devices (grant no. 2022B1212010008), the Fundamental Research Funds for the Central Universities, Sun Yat-sen University (Grant No. 23qnpy57), and the Shenzhen International Quantum Academy. High-pressure synchrotron X-ray measurements were performed at the 4W2 High-Pressure Station, Beijing Synchrotron Radiation Facility, which is supported by the Chinese Academy of Sciences (grant nos. KJCX2-SW-N20 and KJCX2-SW-N03).

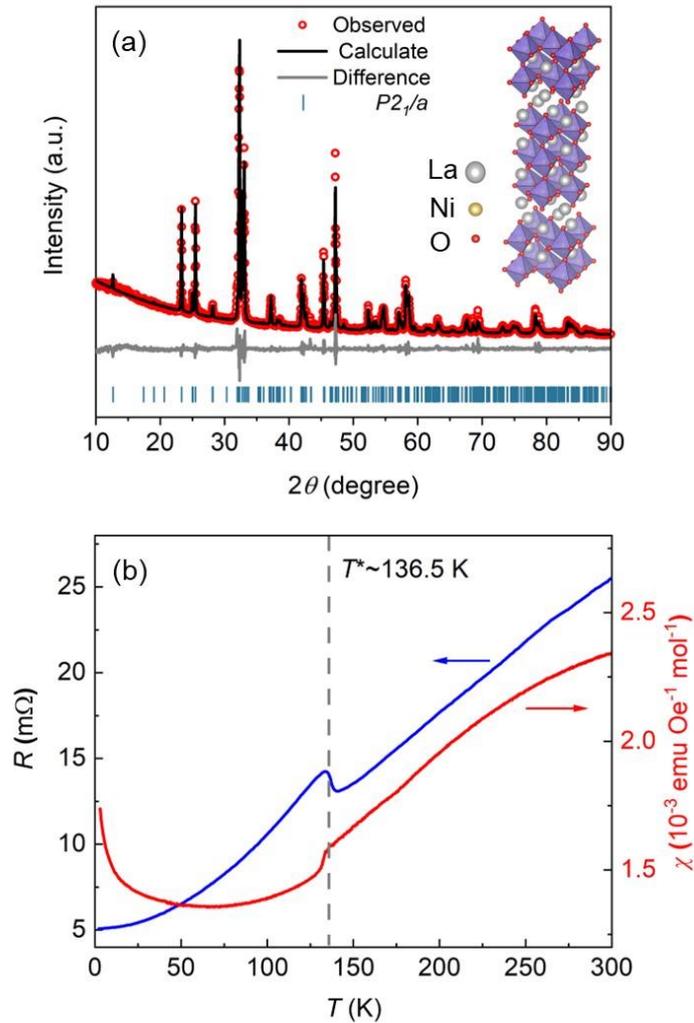

Figure 1 Basic properties of $La_4Ni_3O_{10}$ at ambient pressure. (a) Powder X-ray diffraction pattern and its Rietveld fitting curve of $La_4Ni_3O_{10}$. The inset shows the crystal structure of $La_4Ni_3O_{10}$. Lanthanum, nickel, and oxygen atoms are denoted as grey, yellow, and red dots, respectively. (b) Temperature-dependent resistance (blue line) and zero-field cooling (ZFC) magnetic susceptibility (red line) of $La_4Ni_3O_{10}$ single crystal. The ZFC susceptibility curve is measured with a magnetic field parallel to the c axis. The dashed line marks the anomalies in resistance and susceptibility at 136.5K.

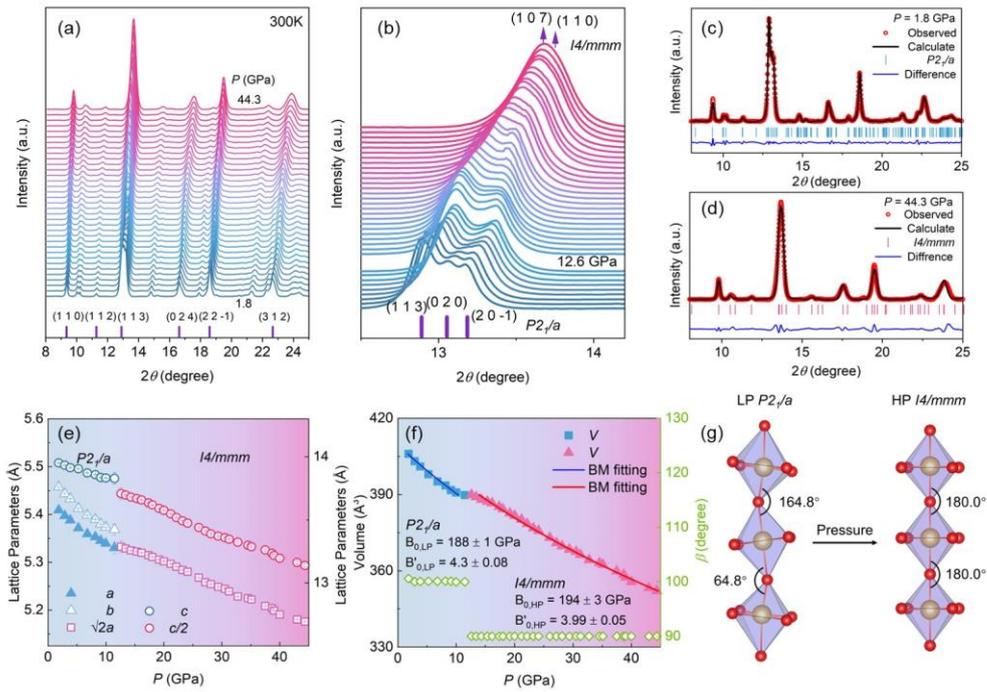

Figure 2 Structural characterizations of pressurized $La_4Ni_3O_{10}$. (a) Synchrotron powder X-ray diffraction patterns of pressurized $La_4Ni_3O_{10}$ up to 44.3 GPa. (b) Details of the evolutions of the diffraction peaks under pressure. (c)-(d) Synchrotron powder X-ray diffraction patterns and their Rietveld fitting curves of $La_4Ni_3O_{10}$ under 1.8 and 44.3 GPa. (e) Evolution of the lattice parameters under pressure. (f) Pressure dependence of the cell volume under pressure. As indicated by the dark blue line and the rose line, the evolution of the cell volume as a function of pressure can be fitted by the Birch-Murnaghan equation. Green blocks denote the variation of the $\beta$ angle against pressure. (g) Enlarged sketches of the trilayer $NiO_6$ octahedra. The Ni-O-Ni angle between two adjacent octahedra layers changes from 164.8° to 180° under pressure.

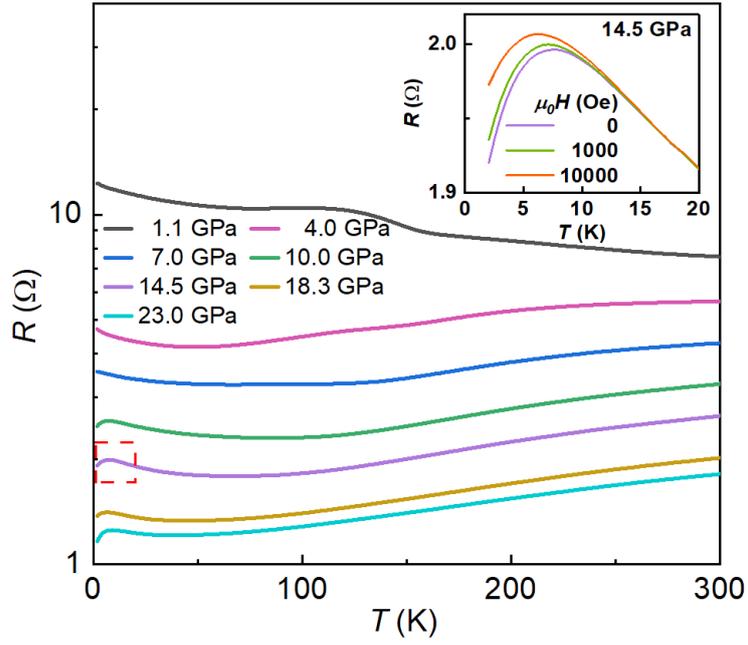

Figure 3 Resistance of pressurized $La_4Ni_3O_{10}$. Temperature-dependent resistance of $La_4Ni_3O_{10}$ under pressure in the range of 1.1 ~ 23.0 GPa. A rapid drop appears in resistance above 10.0 GPa and below 7 K. The inset is the field dependence of the drop at 14.5 GPa as the red dashed square marks.

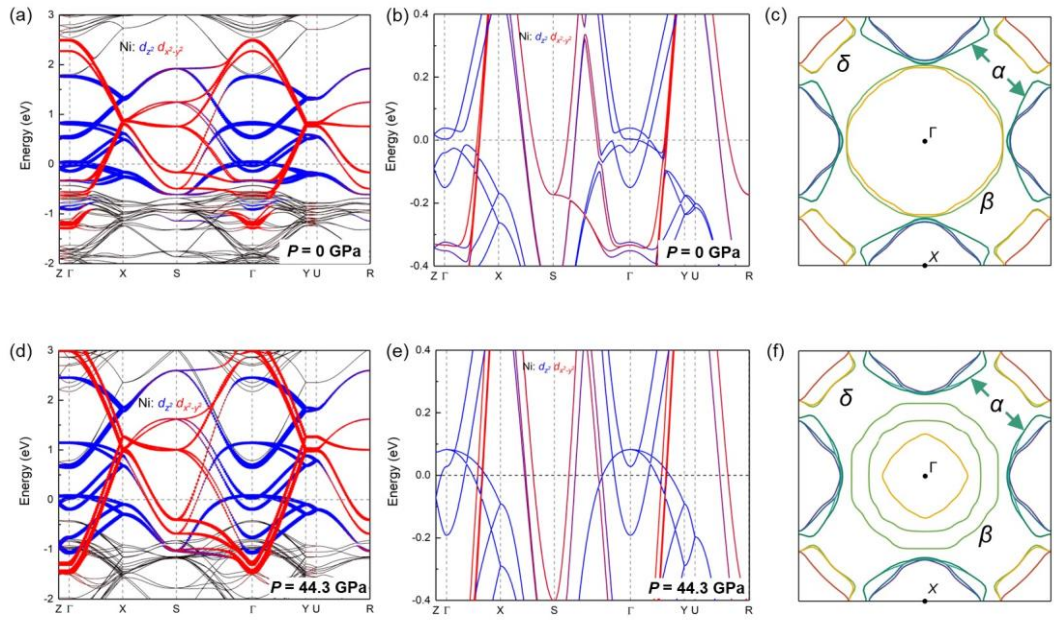

Figure 4 DFT calculations for band structures and Fermi surfaces of $La_4Ni_3O_{10}$ at ambient pressure and 44.3 GPa. (a) Projected band structures of $La_4Ni_3O_{10}$ at ambient pressure. The blue and red lines denote the weight of $3d_{z^2}$ and $3d_{x^2-y^2}$ orbitals of Ni. (b) Enlarged details of the band structures around the Fermi level at ambient. (c) Calculated 2D Fermi surface slice of $La_4Ni_3O_{10}$ at ambient pressure. Different colors represent different bands in the First Brillouin zone at the Fermi level. Hole pockets around the X point of the Brillouin zone are formed by the α bands (denoted by the dark green lines). There are electron pockets centered at the Γ point (the β bands) and electron pockets at the corner (the δ bands) of the Brillouin zone. (d-f) Identical plots of the calculated electronic structures at 44.3 GPa. The splitting of the β band is enhanced by pressure. An additional electron pocket centered at the Γ point crosses the Fermi level, as denoted by the yellow line.

# Supplementary Information "Structural transition, electric transport, and electronic structures in the compressed trilayer nickelate La$_4$Ni$_3$O$_{10}$"

Jingyuan Li[1,*], Cui-Qun Chen[1,*], Chaoxin Huang[1], Yifeng Han[2], Mengwu Huo[1], Xing Huang[1], Peiyue Ma[1], Zhengyang Qiu[1], Junfeng Chen[1], Xunwu Hu[1], Lan Chen[1], Tao Xie[1], Bing Shen[1], Hualei Sun[3,#], Dao-Xin Yao[1,§], and Meng Wang[1,†]

[1]Center for Neutron Science and Technology, Guangdong Provincial Key Laboratory of Magnetoelectric Physics and Devices, School of Physics, Sun Yat-Sen University, Guangzhou, Guangdong 510275, China

[2] Center for Materials of the Universe, School of Molecular Sciences, Arizona State University, Tempe, AZ 85287, USA

[3] School of Sciences, Sun Yat-Sen University, Shenzhen, Guangdong 518107, China

# sunhlei@mail.sysu.edu.cn

§ yaodaox@mail.sysu.edu.cn

† wangmeng5@mail.sysu.edu.cn

The crystal parameters for all measured pressures and the atomic positions at ambient pressure are refined from the experimental data and listed in Table S1 and S2, respectively. For the high-pressure data, the positions of oxygen could not be determined accurately due to the small scattering cross section of O atoms. Thus, the atom positions optimized theoretically are shown in the following TABLE S3-S13. All the atoms have been assumed to be fully occupied.

**TABLE S1** Crystal parameters of pressurized La$_4$Ni$_3$O$_{10}$ obtained from the Rietveld refinements.

| Pressure (GPa) | a (Å) | b (Å) | c (Å) | β (°) | Space group | R$_{wp}$ (%) | R$_p$ (%) | GOF |
|---|---|---|---|---|---|---|---|---|
| 0 | 5.4164(2) | 5.4675(2) | 14.2279(3) | 100.752(4) | P2$_1$/a (Z=2) | 4.72 | 3.43 | 2.24 |
| 2.9 | 5.3974(3) | 5.4434(3) | 13.933(2) | 100.008(14) | P2$_1$/a (Z=2) | 4.02 | 7.51 | 0.52 |
| 5.2 | 5.3714(4) | 5.4125(3) | 13.904(3) | 100.09(2) | P2$_1$/a (Z=2) | 4.98 | 9.05 | 0.66 |
| 8.2 | 5.3529(4) | 5.3894(4) | 13.858(3) | 100.13(2) | P2$_1$/a (Z=2) | 3.60 | 6.81 | 0.49 |
| 10.0 | 5.3393(5) | 5.3729(5) | 13.832(3) | 99.97(3) | P2$_1$/a (Z=2) | 9.96 | 15.55 | 0.49 |
| 15.3 | 3.7640(3) | 3.7640(3) | 27.336(4) | 90 | I4/mmm | 5.66 | 9.15 | 0.76 |
| 19.2 | 3.7522(3) | 3.7522(3) | 27.181(3) | 90 | I4/mmm | 5.01 | 9.07 | 0.69 |
| 26.0 | 3.7256(4) | 3.7256(4) | 26.879(4) | 90 | I4/mmm | 6.11 | 11.32 | 0.85 |
| 30.5 | 3.7090(4) | 3.7090(4) | 26.716(5) | 90 | I4/mmm | 6.07 | 10.97 | 0.82 |
| 35.1 | 3.6932(5) | 3.6932(5) | 26.574(5) | 90 | I4/mmm | 6.56 | 12.16 | 0.87 |
| 40.0 | 3.6710(7) | 3.6710(7) | 26.391(8) | 90 | I4/mmm | 8.25 | 14.43 | 0.95 |
| 44.3 | 3.6606(7) | 3.6606(7) | 26.277(8) | 90 | I4/mmm | 8.16 | 14.16 | 1.06 |

TABLE S2 Refined atomic positions for La$_4$Ni$_3$O$_{10}$ at ambient pressure ($P2_1/a$).

| Name | x | y | z |
|---|---|---|---|
| La1 | -0.195(3) | 0.511(2) | -0.8975(1) |
| La2 | -0.060(2) | 0.499(2) | -0.6380(2) |
| Ni1 | 0 | 0 | -0.5 |
| Ni2 | -0.1431(7) | -0.009(8) | -0.773(4) |
| O1 | -0.218(3) | -0.017(2) | -0.927(2) |
| O2 | 0.079(2) | -0.042(2) | -0.355(2) |
| O3 | 0.119(3) | -0.257(2) | -0.764(3) |
| O4 | 0.108(3) | 0.245(2) | -0.788(2) |
| O5 | -0.254(9) | -0.261(3) | -0.514(2) |

TABLE S3 Optimized atomic positions for La$_4$Ni$_3$O$_{10}$ at 2.9 GPa ($P2_1/a$).

| Name | x | y | z |
|---|---|---|---|
| La1 | -0.19756 | 0.49067 | -0.8955 |
| La2 | -0.05497 | 0.50198 | -0.63428 |
| Ni1 | 0 | 0 | -0.5 |
| Ni2 | -0.1383 | -0.0017 | -0.77902 |
| O1 | -0.21256 | -0.02989 | -0.92982 |
| O2 | 0.07268 | -0.04391 | -0.3592 |
| O3 | 0.11617 | -0.25013 | -0.7695 |
| O4 | 0.1056 | 0.25141 | -0.7886 |
| O5 | -0.23654 | -0.26869 | -0.51132 |

TABLE S4 Optimized atomic positions for La$_4$Ni$_3$O$_{10}$ at 5.2 GPa ($P2_1/a$).

| Name | x | y | z |
|---|---|---|---|
| La1 | -0.19898 | 0.49204 | -0.89573 |
| La2 | -0.077 | 0.50139 | -0.63493 |
| Ni1 | 0 | 0 | -0.5 |
| Ni2 | -0.14147 | -0.00149 | -0.77922 |
| O1 | -0.21478 | -0.0258 | -0.93006 |
| O2 | 0.07085 | -0.03923 | -0.35908 |
| O3 | 0.11477 | -0.24784 | -0.77052 |
| O4 | 0.10166 | 0.25305 | -0.78731 |
| O5 | -0.28006 | -0.22539 | -0.5105 |

TABLE S5 Optimized atomic positions for La$_4$Ni$_3$O$_{10}$ at 8.2 GPa ($P2_1/a$).

| Name | x | y | z |
|---|---|---|---|
| La1 | -0.19878 | 0.49303 | -0.89575 |
| La2 | -0.07478 | 0.50138 | -0.63482 |
| Ni1 | 0 | 0 | -0.5 |
| Ni2 | -0.1413 | -0.00125 | -0.77942 |
| O1 | -0.21467 | -0.02267 | -0.93016 |
| O2 | 0.07135 | -0.03558 | -0.35901 |
| O3 | 0.11467 | -0.24795 | -0.77107 |
| O4 | 0.10252 | 0.25297 | -0.78621 |
| O5 | -0.27699 | -0.2279 | -0.50952 |

**TABLE S6** Optimized atomic positions for $La_4Ni_3O_{10}$ at 10.0 GPa ($P2_1/a$).

| Name | x | y | z |
|---|---|---|---|
| La1 | -0.19864 | 0.49378 | -0.89577 |
| La2 | 0.07357 | 0.50131 | -0.63477 |
| Ni1 | 0 | 0 | -0.5 |
| Ni2 | -0.14117 | -0.00108 | -0.77955 |
| O1 | -0.21461 | -0.02033 | -0.93025 |
| O2 | 0.07159 | -0.03244 | -0.35896 |
| O3 | 0.11464 | -0.24796 | -0.77156 |
| O4 | 0.10313 | 0.25296 | -0.78533 |
| O5 | -0.27542 | -0.22903 | -0.50871 |

**TABLE S7** Optimized atomic positions for $La_4Ni_3O_{10}$ at 15.3 GPa ($I4/mmm$).

| Name | x | y | z |
|---|---|---|---|
| La1 | 0 | 0 | 0.30221 |
| La2 | 0 | 0 | 0.43316 |
| Ni1 | 0 | 0 | 0 |
| Ni2 | 0 | 0 | 0.13975 |
| O1 | 0 | 0.5 | 0 |
| O2 | 0 | 0 | 0.07061 |
| O3 | 0 | 0.5 | 0.13885 |
| O4 | 0 | 0 | 0.21534 |

**TABLE S8** Optimized atomic positions for $La_4Ni_3O_{10}$ at 19.2 GPa ($I4/mmm$).

| Name | x | y | z |
|---|---|---|---|
| La1 | 0 | 0 | 0.30218 |
| La2 | 0 | 0 | 0.43309 |
| Ni1 | 0 | 0 | 0 |
| Ni2 | 0 | 0 | 0.13993 |
| O1 | 0 | 0.5 | 0 |
| O2 | 0 | 0 | 0.07063 |
| O3 | 0 | 0.5 | 0.13877 |
| O4 | 0 | 0 | 0.21537 |

**TABLE S9** Optimized atomic positions for $La_4Ni_3O_{10}$ at 26.0 GPa ($I4/mmm$).

| Name | x | y | z |
|---|---|---|---|

| Name | x | y | z |
| --- | --- | --- | --- |
| La1 | 0 | 0 | 0.30211 |
| La2 | 0 | 0 | 0.43297 |
| Ni1 | 0 | 0 | 0 |
| Ni2 | 0 | 0 | 0.14025 |
| O1 | 0 | 0.5 | 0 |
| O2 | 0 | 0 | 0.07068 |
| O3 | 0 | 0.5 | 0.13858 |
| O4 | 0 | 0 | 0.21543 |

**TABLE S10** Optimized atomic positions for La$_4$Ni$_3$O$_{10}$ at 30.5 GPa (*I4/mmm*).

| Name | x | y | z |
| --- | --- | --- | --- |
| La1 | 0 | 0 | 0.30207 |
| La2 | 0 | 0 | 0.43291 |
| Ni1 | 0 | 0 | 0 |
| Ni2 | 0 | 0 | 0.14041 |
| O1 | 0 | 0.5 | 0 |
| O2 | 0 | 0 | 0.07071 |
| O3 | 0 | 0.5 | 0.13846 |
| O4 | 0 | 0 | 0.21548 |

**TABLE S11** Optimized atomic positions for La$_4$Ni$_3$O$_{10}$ at 35.1 GPa (*I4/mmm*).

| Name | x | y | z |
| --- | --- | --- | --- |
| La1 | 0 | 0 | 0.30204 |
| La2 | 0 | 0 | 0.43286 |
| Ni1 | 0 | 0 | 0 |
| Ni2 | 0 | 0 | 0.14054 |
| O1 | 0 | 0.5 | 0 |
| O2 | 0 | 0 | 0.07073 |
| O3 | 0 | 0.5 | 0.13835 |
| O4 | 0 | 0 | 0.21552 |

**TABLE S12** Optimized atomic positions for La$_4$Ni$_3$O$_{10}$ at 40.0 GPa (*I4/mmm*).

| Name | x | y | z |
| --- | --- | --- | --- |
| La1 | 0 | 0 | 0.30199 |
| La2 | 0 | 0 | 0.43280 |
| Ni1 | 0 | 0 | 0 |
| Ni2 | 0 | 0 | 0.14071 |
| O1 | 0 | 0.5 | 0 |
| O2 | 0 | 0 | 0.07076 |
| O3 | 0 | 0.5 | 0.13819 |
| O4 | 0 | 0 | 0.21558 |

**TABLE S13** Optimized atomic positions for $La_4Ni_3O_{10}$ at 44.3 GPa ($I4/mmm$).

| Name | x | y | z |
|---|---|---|---|
| La1 | 0 | 0 | 0.30198 |
| La2 | 0 | 0 | 0.43277 |
| Ni1 | 0 | 0 | 0 |
| Ni2 | 0 | 0 | 0.14082 |
| O1 | 0 | 0.5 | 0 |
| O2 | 0 | 0 | 0.07078 |
| O3 | 0 | 0.5 | 0.13812 |
| O4 | 0 | 0 | 0.21561 |